\input{aipcheck}

\documentclass[
    ,final            
  ]
  {aipproc}

\layoutstyle{8x11single}

\begin{document}

\title{Non-perturbative Jet Quenching from Geometric Data}


\author{Jinfeng Liao}{
  address={Physics Department, Bldg. 510A,
   Brookhaven National Laboratory, Upton, NY 11973, USA}
}

\begin{abstract}
 This contribution discusses the geometric tomography by highly energetic
 jets penetrating the hot QCD matter in heavy ion collisions from RHIC to
 LHC energies. In particular the geometric data on the azimuthal anisotropy of high $p_t$ hadrons discriminates different models
 and strongly hints at energy loss mechanism beyond those based on perturbation theory. Taking together the RHIC and LHC data, the comparison with models is in favor of the model with strong enhancement of jet quenching in near-$T_c$ matter.
\end{abstract}

\maketitle


\section{Introduction}

Highly energetic jets born from initial hard collisions
provide natural ``tomography'' of the hot QCD matter created
in a heavy ion collision. Jet quenching due to energy loss along
the jet path through the medium encodes essential information about
the dynamics of jet-medium interaction and the medium properties as well,
which shall be inferrable from experimental observables such as high
$p_t$ hadron suppression and di-hadron correlations (for reviews see e.g. \cite{Gyulassy:2003mc}).
While the jet quenching has been experimentally established as a very robust
 phenomenon, the microscopic mechanism of energy loss is not yet fully understood. The geometric features of jet quenching observables are particularly useful in discriminating different models of energy loss. These include:
\begin{description}
\item[A-dependence:]how the jet energy loss depends on the size of the colliding systems(e.g. AuAu v.s. CuCu);
\item[b-dependence:]how the jet energy loss changes with the medium ``thickness'' at varied collision impact parameter;
\item[$\phi$-dependence:] how the jet energy loss is related to its azimuthal angle $\phi$ with respect to the reaction plane.
\end{description}
For any given dynamical or phenomenological model, one can fix the model parameters by fitting data in the most central collisions and then ``predict'' the above geometric dependence as a crucial test of the model.

Let's focus on the $\phi$-dependence. In non-central collisions, the medium ``thickness'' as seen by a penetrating jet depends on the azimuthal angle of the jet with respect to the reaction plane therefore leading to the reaction-plane dependence of high-$p_t$ hadron suppression i.e. $R_{AA}(\phi)$ \cite{Gyulassy:2000gk}. This is directly related to the azimuthal anisotropy parameter for high-$p_t$ hadrons, $V_2^{hard}$ . Despite the success of many models in describing the overall ``opacity'' or nuclear modification factor $R_{AA}$ and its centrality dependence, it was known for long time that almost all of those models significantly under-predicted the $V_2^{hard}$ and failed the test by geometric data \cite{Shuryak:2001me}\cite{Adler:2006bw}. The lack of a simultaneous description for $R_{AA}$ and $V_2^{hard}$ in a single model was not resolved till a new insight suggested in \cite{Liao:2008dk}. Motivated by the ``magnetic scenario'' for sQGP \cite{Liao:2006ry}, the authors of \cite{Liao:2008dk} pointed that the energy loss of a jet may not simply scale with the local medium density as most models have assumed, but actually have nontrivial dependence on matter density (or temperature). It was particularly shown that including a jet quenching component with strong enhancement in the near-$T_c$ matter successfully explains the geometric data for the first time. Such an enhancement of jet-medium interaction may originate from  non-perturbative structures created by the (color-)electric jet passing a plasma of (color-)magnetic monopoles that dominate the near-$T_c$ matter \cite{Liao:2006ry},\cite{Liao:2008vj}. More recently there appeared another class of jet quenching models with a much stronger path-length dependence of energy loss $\Delta E \sim L^3$ than the usual $L^2$ dependence from LPM effect for multiple gluon radiation, which also managed to describe the $R_{AA}$ and $V_2^{hard}$ data at the same time \cite{Marquet:2009eq}\cite{Jia:2010ee}. The strong path-length dependence in these models was motivated by AdS/CFT calculations for certain strongly coupled Yang-Mills plasma. The two different models both generate large $V_2^{hard}$ (for fixed $R_{AA}$) because they both enhance the energy loss in the outer-layer of fireball where the eccentricity is larger \cite{Jia:2011pi}. We therefore have seen that the geometric data at RHIC can be explained only by models incorporating non-perturbative jet quenching mechanism.

In this contribution, we will discuss geometric models of jet quenching without and with (varied) non-perturbative elements by confronting them with high-precision RHIC data as well as by comparison with preliminary LHC data.

\begin{figure} \label{fig_RHIC}
  \includegraphics[height=.275\textheight]{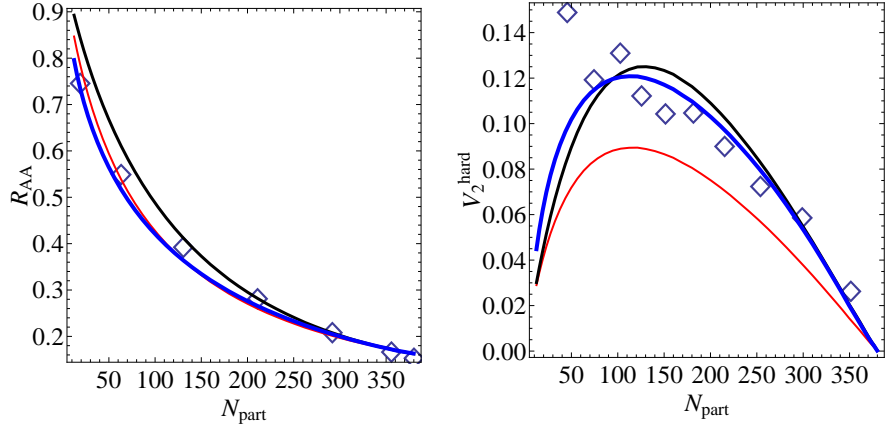}
  \caption{High-$p_t$ hadron $R_{AA}$(left) and $V_2$(right) versus $N_{part}$: a comparison between RHIC data and calculations from $L^2$ model(red), near-$T_c$ enhancement model(blue), and $L^3$ model(black) respectively (see text for details).}
\end{figure}

\section{Geometric Models and Comparison with Data}

Geometric models for jet quenching, though lacking many dynamical details,
reflect the general geometric features (e.g. the path-length dependence) that are most crucial for describing geometric data \cite{Shuryak:2001me}\cite{Liao:2008dk}\cite{Jia:2010ee}\cite{Jia:2011pi}. In such models the final energy $E_f$ of a jet with initial energy $E_i$ after traveling an in-medium path $P$ can be parameterized as $E_f = E_i \times f_P$ with the suppression factor $f_P$ given by:
\begin{equation}
f_P = exp\left\{ - \int_P\, \kappa[s(l)]\, s(l)\, l^m dl  \right\}
\end{equation}
In the above the $s(l)$ is the entropy density of local matter at a given point on the jet path, while the $\kappa(s)$ is the local jet quenching strength which as a property of matter should in principle depend on the local density $s(l)$. After averaging over all jet paths (including all possible start points and orientations) one may obtain the nuclear modification factor:
\begin{equation}
R_{AA} = <\, (f_P)^{n-2}  \,>_P
\end{equation}
where the exponent $n$ comes from measured reference p-p spectrum (see e.g. \cite{Adler:2006bw} for a detailed account). Alternatively one may also study the reaction-plane dependence by averaging over jet paths with a particular azimuthal orientation, i.e. $R_{AA}(\phi) = <\, (f_P)^{n-2}  \,>_{P(\phi)}$ from which $V_2^{hard}$ is derivable. We study three classes of models here:
\begin{description}
\item[$L^2$ model:] assuming $m=1$ (i.e. square path-length dependence as per LPM) and $\kappa(s)=\kappa$ as a constant (i.e. energy loss simply proportional to local density) which are common features of most jet energy loss models;
\item[near-$T_c$ enhancement model:] also assuming $m=1$ but introducing a strong jet quenching component in the vicinity of $T_c$ (with density $s_c$ and span of $s_w$) via $\kappa(s)=\kappa [1+ \xi\, exp(-(s-s_c)^2/s_w^2)]$ with $\xi=6$ (see \cite{Liao:2008dk} for details);
\item[$L^3$ model:] assuming $m=2$ (i.e. cubic path-length dependence) while keeping $\kappa(s)=\kappa$ as a constant.
\end{description}
After fixing the parameter of each in the most central collisions at RHIC 200GeV, one can then compare the predictions for $R_{AA}$ and $V_2^{hard}$ at different centralities from each model with PHENIX data\cite{Adare:2010sp}: see Fig.\ref{fig_RHIC}. While all three describe $R_{AA}$ very well, the $L^2$ model generates too little $V_2^{hard}$ and only the near-$T_c$ enhancement model and the $L^3$ model can account for the sizeable  anisotropy. Therefore we emphasize again that the geometric data of jet quenching at RHIC strongly favor models with certain non-perturbative mechanisms.

\begin{figure} \label{fig_LHC}
  \includegraphics[height=.275\textheight]{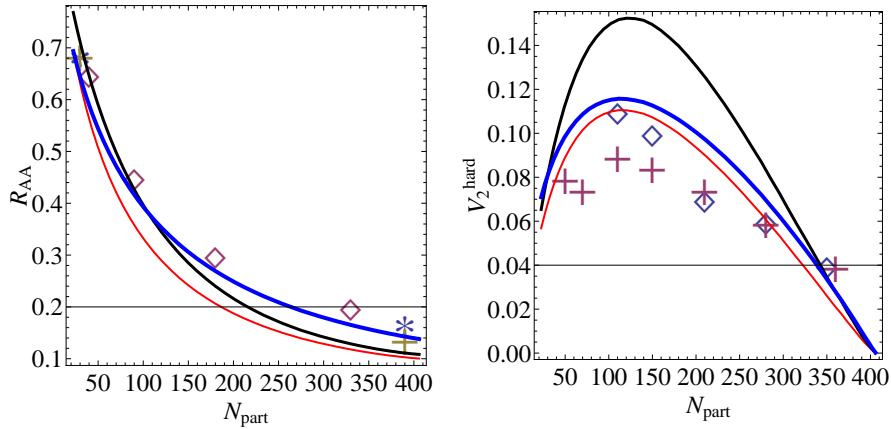}
  \caption{High-$p_t$ hadron $R_{AA}$(left) and $V_2$(right) versus $N_{part}$: a comparison between preliminary LHC data and calculations from $L^2$ model(red), near-$T_c$ enhancement model(blue), and $L^3$ model(black) respectively (see text for details).}
\end{figure}

It is natural to ask whether the last two models could be further distinguished: this could be answered by applying the models (calibrated at RHIC) to LHC collisions and comparing with data. In Fig.\ref{fig_LHC} we present for the first time such a comparison with very preliminary LHC data as extracted from plots in pertinent experimental talks by ATLAS and ALICE at Quark Matter 2011 \cite{QM11}. From the figure one can see that:1) both $L^2$ and $L^3$ models somewhat over-quench the jets while the near-$T_c$ model gives a fairly good description of $R_{AA}$; 2) the $L^3$ model continues to predict a strong anisotropy overshooting the data while the $V_2^{hard}$ from $L^2$ and near-$T_c$ models are in reasonable agreement with data.


\section{Conclusion}

To conclude, by studying geometric models of jet quenching at both RHIC and LHC energies, we have shown that the model assuming a non-perturbative component with strongly enhanced jet quenching in near-$T_c$ matter are best supported by the geometric data. We point out in passing that such a scenario, featuring non-monotonic dependence of transport properties on matter density/temperature near the phase boundary, has many supportive evidences from various other studies on jet quenching, fragmentation, heavy quark, viscosity and energy loss relation, etc\cite{misc}.


\begin{theacknowledgments}
 The author is grateful to R. Fries, M. Gyulassy, W. Horowitz, V. Koch, J. Jia, P. Jocobs, H. Pirner, R. Rapp, X. Wang, and F. Yuan for helpful discussions on various aspects related to this contribution. He also thanks the organizers for generous conference support. The research is supported under DOE Contract No. DE-AC02-98CH10886.
\end{theacknowledgments}



\bibliographystyle{aipproc}   





\end{document}